\documentstyle[prl,aps,psfig]{revtex}
\newcommand{\be}{\begin{eqnarray}}
\newcommand{\ee}{\end{eqnarray}}

\begin{document}
\twocolumn[\hsize\textwidth\columnwidth\hsize
           \csname @twocolumnfalse\endcsname
\title{General response function for interacting quantum liquids}
\author{Klaus Morawetz$^{1,2}$, Uwe Fuhrmann$^{2}$}
\address{$^1$ LPC-ISMRA, Bld Marechal Juin, 14050 Caen
 and  GANIL, Bld Becquerel, 14076 Caen Cedex 5, France\\
$^2$ Fachbereich Physik, University Rostock, D-18055 Rostock,Germany}
\maketitle
\begin{abstract}
Linearizing the appropriate kinetic equation we derive 
general response functions including selfconsistent mean fields or
density functionals and collisional dissipative contributions. The latter ones are considered in relaxation time approximation conserving successively different balance equations.
The effect of collisions is represented by correlation functions which
are possible to calculate with the help of the finite temperature
Lindhard RPA expression. The presented results are applicable to
finite temperature response of interacting quantum systems if the
quasiparticle or mean field energy is parameterized within Skyrme -
type of functionals including density, current and energy
dependencies which can be considered alternatively as density functionals. 
By this way we allow to share correlations between density functional
and collisional dissipative contributions appropriate for the special treatment.
We present results for collective modes like the plasmon in plasma
systems and the giant resonance in nuclei. The collisions lead in general
to an enhanced damping of collective modes. If the collision frequency
is close to the frequency of the collective mode, resonance occurs and
the collective mode is enhanced showing a collisional narrowing.
\end{abstract}
\vskip2pc]

\section{Introduction}

The response of matter to an external perturbation is the main source of knowledge about the matter itself. For instance, in plasma physics the polarization function $\Pi(q,\omega)$ is linked via the dielectric function $\epsilon(q,\omega)$ to the electrical conductivity $\sigma(q,\omega)$ by
\be
\epsilon(q,\omega)=1-V(q) \Pi(q,\omega)=1+{i\over \omega} \sigma(q,\omega).
\label{eps}
\ee
In nuclear matter, e.g., the response function $\chi=\Pi/\epsilon$ allows one to study excitations and giant resonances which in turn yields information about the equation of state like the isothermal compressibility which is given by
\be
&&\kappa={1\over n^2}\left ({\partial n\over \partial \mu}\right )_T={1\over n^2 T}\lim\limits_{q\to0}\int {d\omega\over \pi} {1\over {\rm e}^{\omega/T}-1} {\rm Im}{\chi(q,\omega)}
\nonumber\\&&
\label{kap}
\ee
or to calculate fluctuations and diffusion coefficients.

Two lines of theoretical improvements of the response function can be found in print recently. The first one starts from TDHF equations and considers the response of nuclear matter described by a time reversal broken Skyrme interaction \cite{BVA95,HNP96}. The other line tries to improve the response by the inclusion of collisional correlations \cite{Mer70,HPR93,KH94,RW98} and for multicomponent systems \cite{MWF97,Br99}. In this paper we want to combine both lines of improvements into one 
expression and derive therefore the response function from a kinetic equation including mean field (Skyrme) and collisional correlations.

We consider here interacting matter which can be described by an
energy functional ( mean field) ${\cal E}$ originally introduced by
Skyrme \cite{S56,S59} and the residual interaction. The latter one we
condense in a collisional integral additional to the TDHF
equation. Then the response to an external perturbation will contain
the effect of Skyrme mean field and additionally the effect of residual
interaction. While this schema and the results are of general interest
for any interacting Fermi- or Bose system, we will only mention
application examples from nuclear matter and plasma physics. For the
latter one we might consider the energy functional as a
parameterization of the selfenergy in line with the philosophy of
density functional theory. By this way we have the freedom to share
the correlations between mean field like density functional
parameterizations and explicit collisional or dissipative like
correlations which are condensed
in a relaxation time. Of course, when deriving this parameterizations
microscopically special care is required to avoid double counting of correlations.

Specifically, we want to obtain the density, current and energy response $\chi, \chi_J, \chi_E$ of an interacting quantum system 
\be
&&\left (\matrix {\delta n\cr\delta {\bf \nabla J}\cr \delta E} \right )=\left (\matrix{\chi \cr \chi_J \cr \chi_E}\right ) \,\,V^{\rm ext}\equiv{\cal X} \left (\matrix{1\cr 0\cr 0}\right ) V^{\rm ext}\equiv{\cal X}\nu^{\rm ext}
\nonumber\\
&&
\label{def}
\ee
to the external perturbation $V^{\rm ext}$ provided the density, momentum and energy are conserved
\be
n({\bf R},t)&=&\sum \limits_j \langle \Phi_j^*({\bf R})\Phi_j({\bf R})\rangle
=s \sum\limits_p f({\bf p,R},t),\nonumber\\
{\bf J}({\bf R},t)&=&\sum \limits_j \langle {\nabla_{\bf R}-\nabla_{\bf R'}\over 2 i} \Phi_j^*({\bf R'})\Phi_j({\bf R})\rangle_{{\bf R}={\bf R'}}\nonumber\\
&=&
s\sum\limits_p {\bf p} f({\bf p,R},t),\nonumber\\
E({\bf R},t)&=&\sum \limits_j \langle 
{\cal H}({\bf R},t)
\rangle
=s\sum\limits_p \varepsilon({\bf p,R},t) f({\bf p,R},t).
\ee
Here $s$ is the spin-isospin degeneracy and we express the observables in terms of the Wigner function $f({\bf p,R},t)$ which is related to the one-particle density operator $\hat \rho$ by
\be
f({\bf p,R},t)=\sum\limits_q {\rm e}^{i{\bf q R}} \langle p+{q\over 2}|\hat \rho|p-{q\over 2}\rangle
\ee
and introduce the quasiparticle (Skyrme) energy $\varepsilon$ 
\be
\varepsilon({\bf p,R},t)=\sum\limits_q {\rm e}^{i{\bf q R}} \langle p+{q\over 2}|\hat {\cal E}|p-{q\over 2}\rangle
\ee
in the spirit of Landau theory $\varepsilon={\delta E\over \delta
  f}$. We will neglect the contributions of energy gain which arise
from noninstantaneous collisions \cite{LSM99}. Here the energy
functional or mean field (Skyrme) energy $\varepsilon$ is assumed to be parameterized as \cite{EBGKV75}
\be
\hat {\cal E}&=&- \nabla ({1\over 2 m}+\epsilon_1 n)\nabla+\epsilon_2 n+\epsilon_1 [2 m E-{1\over i} (\nabla {\bf J}+{\bf J}\nabla)]\nonumber\\
&&+\epsilon_3\nabla^2 n +\epsilon_4 n^{\alpha+1}.
\label{pa}
\ee
Please note that the occurrence of current contributions $\sim {\bf
  J}$ breaks explicitly the time invariance. These terms appear with
the same coefficient $\epsilon_1$ as the effective mass and energy
contribution in order to ensure Galilean invariance. The density
dependence $\alpha\ne 1$ deviating from the one arising by Skyrme
three-body contact interaction has been introduced and compared with
experiments in \cite{Ko75}.

\section{Derivation of general response function}

We start the derivation of the response from the quantum kinetic equation for the density operator in relaxation time approximation \footnote{
The quasiclassical Landau equation follows from gradient expansion as
\be
{\partial \over \partial t} f+\partial_{\bf p} \epsilon \partial_{\bf r} f-\partial_{\bf r} \epsilon \partial_{\bf p} f=-{f-f^{\rm l.e.}\over \tau}
\ee
}
\be
\dot {\hat \rho}+i[\hat {\cal E}+\hat V^{\rm ext},\hat \rho] ={\hat \rho^{\rm l.e.}-\hat \rho \over \tau}
\label{1}
\ee
where the relaxation is considered with respect to the local density operator $\hat \rho^{\rm l.e.}$ or the corresponding local equilibrium distribution function
\be
f^{\rm l.e.}({\bf p,R},t)=f_0\left ({\varepsilon_0({\bf p}-{\bf Q}({\bf R},t))-\mu({\bf R},t)\over T({\bf R},t)}\right )
\label{2}
\ee
with the (Fermi/Bose) distribution $f_0$.
This local equilibrium is given by a local chemical potential $\mu$,
a local temperature $T$ and a local mass motion momentum $Q$. These
local quantities will be specified by the requirement that the expectation values for density, momentum and energy are the same as the expectation values performed with $f$. 

\subsection{Conservation laws}
From (\ref{1}) we see that the conservation laws for density, momentum and energy are fulfilled if the corresponding expectation value of the collision side vanishes
\be
\sum\limits_p (f-f^{\rm l.e.})&=&0\nonumber\\
\sum\limits_p {\bf p} (f-f^{\rm l.e.})&=&0\nonumber\\
\sum\limits_p \varepsilon (f-f^{\rm l.e.})&=&0.
\ee
Taking this into account we can express the deviation of the observables $\phi=1,{\bf p}, \varepsilon$ from equilibrium
considering $\delta f =f-f_0=f-f^{\rm l.e.}+f^{\rm l.e.}-f_0$ as
\be
\delta \phi({\bf q}, \omega)&=&\sum\limits_p \phi \delta f({\bf p,q},\omega)\nonumber\\
&=&\sum\limits_p \phi (f^{\rm l.e.}-f_0)\nonumber\\
&=&\sum\limits_p \phi {f_0({\bf p}+{{\bf q}\over 2})-f_0({\bf p}-{{\bf q}\over 2}) \over \varepsilon_0({\bf p}+{{\bf q}\over 2})-\varepsilon_0({\bf p}-{{\bf q}\over 2})} \nonumber\\
&&\times\left [-\delta \mu -{\bf q} {\partial \varepsilon_0 \over \partial {\bf p}}\delta {Q}-{\varepsilon_0-\mu \over T} \delta T\right ]
\label{6}
\ee
where we have performed Fourier transform $t\rightarrow -i\omega$ and ${\bf r}\rightarrow i {\bf q}$. In the last line we restrict to {\it linear response} of (\ref{2}). 
We assume for simplicity a homogeneous equilibrium $f_0(\varepsilon_0({\bf p}))$ such that only the deviations $\delta \phi({\bf r},t)$ and $\delta f({\bf p,r},t)$ are space dependent. This is no principle restriction but otherwise a lot of later algebraic expressions would take the form of integral equations. Further $\delta {\bf Q}({\bf q},t)={\bf q} \delta Q({\bf q},t)$ is employed.
With the abbreviation
\be
a_{\phi}&=&\sum\limits_p \phi {f_0({\bf p}+{{\bf q}\over 2})-f_0({\bf p}-{{\bf q}\over 2}) \over \varepsilon_0({\bf p}+{{\bf q}\over 2})-\varepsilon_0({\bf p}-{{\bf q}\over 2})}\nonumber\\
&=&g_\phi(0),\nonumber\\
{b}_{\phi}&=&\sum\limits_p \phi {\bf q} {\partial \epsilon \over \partial {\bf p}}
{f_0({\bf p}+{{\bf q}\over 2})-f_0({\bf p}-{{\bf q}\over 2}) \over \varepsilon_0({\bf p}+{{\bf q}\over 2})-\varepsilon_0({\bf p}-{{\bf q}\over 2})}\nonumber\\
&=&g_{\phi {\bf q} \partial_{\bf p} \varepsilon_0}(0),
\nonumber\\
c_{\phi}&=&\sum\limits_p \phi {\varepsilon_0-\mu \over T} {f_0({\bf p}+{{\bf q}\over 2})-f_0({\bf p}-{{\bf q}\over 2}) \over \varepsilon_0({\bf p}+{{\bf q}\over 2})-\varepsilon_0({\bf p}-{{\bf q}\over 2})}
\nonumber\\
&=&{1 \over T} g_{\phi \varepsilon_0}(0)-{\mu \over T}g_{\phi}(0) 
\label{abb}
\ee
and the correlation function
\be
g_{\phi}(\omega)&=&\sum\limits_p \phi {f_0({\bf p}+{{\bf q}\over 2})-f_0({\bf p}-{{\bf q}\over 2}) \over \varepsilon_0({\bf p}+{{\bf q}\over 2})-\varepsilon_0({\bf p}-{{\bf q}\over 2})-\omega-i0}
\label{g}
\ee
we can write the deviation of the observables from equilibrium
according to (\ref{6}) explicitly
\be
\delta n&=&-\delta\mu a_1-\delta Q b_1 -\delta T c_1, \nonumber\\
\delta J_q&=&{\bf q} \delta {\bf J}=-\delta\mu a_{\bf q p}-\delta {Q} b_{\bf q p} -\delta T c_{\bf q p}, \nonumber\\
\delta E&=&-\delta\mu a_{\epsilon}-\delta Q b_{\epsilon} -\delta T c_{\epsilon}.
\label{f1}
\ee

Instead of the vector equation for the current ${\bf J}$ we consider the
projection onto the direction of ${\bf q}$. This simplifies matters as long
as we have no active media and ${\bf J} || {\bf Q}$.

\subsection{Response from kinetic equation}
To derive the response function  we will obtain a second equation set from linearizing the kinetic equation (\ref{1}) and the corresponding balance equations. Fourier transform $t\rightarrow -i\omega$ and ${\bf r}\rightarrow i {\bf q}$ the equation (\ref{1}) can be linearized
\be
&&-i \omega \delta f+i \left [\varepsilon_0({\bf p}+{{\bf q}\over 2})-\varepsilon_0({\bf p}-{{\bf q}\over 2})\right ]\delta f\nonumber\\
&&-i \left [f_0({\bf p}+{{\bf q}\over 2})-f_0({\bf p}-{{\bf q}\over 2})\right ]
\nonumber\\&&\times
\left [V^{\rm ext}+(V_0 +V_4 p^2) \delta n + {\bf p q} V_1 \delta J_q+V_2 \delta E \right ]
\nonumber\\
&&=-{\delta f\over \tau }
\nonumber\\&&
+{1\over \tau}{f_0({\bf p}+{{\bf q}\over 2})-f_0({\bf p}-{{\bf q}\over 2}) \over \varepsilon_0({\bf p}+{{\bf q}\over 2})-\varepsilon_0({\bf p}-{{\bf q}\over 2})} \left [-\delta \mu -{\bf q} {\partial \varepsilon_0 \over \partial {\bf p}} \delta Q-{\varepsilon_0-\mu \over T} \delta T \right ].\nonumber\\&&
\label{l1}
\ee
In (\ref{l1}) the mean field contributions $V_i$ will lead just to selfconsistency.
The coefficients $V_i$ it-selves are linked to the parameterization (\ref{pa}) as
\be
V_0&=&{\delta \varepsilon_0 \over \delta n}=\epsilon_2-\epsilon_1 {q^2\over 2}-\epsilon_3 q^2 +(\alpha+1) n_0^{\alpha} \epsilon_4,\nonumber\\
V_1&=& {\delta \varepsilon_0 \over \delta {J_q}}=-{2 \epsilon_1\over q^2},\nonumber\\
V_2&=&{\delta \varepsilon_0 \over \delta E}=2 m \epsilon_1,\nonumber\\
V_4&=&\epsilon_1.
\label{mf}
\ee

We can solve (\ref{l1}) for $\delta f$ and perform momentum integrations to obtain the observables $\delta n$, $\delta J_q$, $\delta E$. This leads to the following closed equation system
\be
{\cal I} \left (\matrix{\delta n\cr\delta J_q\cr \delta E}\right )={\cal B}\left (\matrix{\delta \mu \cr\delta Q\cr \delta T}\right )+
{\cal V} \left (\matrix{\delta n\cr\delta J_q\cr \delta E}\right )+\left (\matrix{g_1\cr g_{\bf p q}\cr  g_\epsilon}\right ) V^{\rm ext}
\label{g1}
\ee
together with the set (\ref{f1})
\be
 \left (\matrix{\delta n\cr\delta J_q\cr \delta E}\right )={\cal A}\left (\matrix{\delta \mu \cr\delta Q \cr \delta T}\right ).
\label{g2}
\ee
The matrices are
\be
{\cal V}&=&\left (\matrix{g_1 V_0 +g_{p^2} V_4&g_{\bf pq} V_1 &g_1 V_2\cr
g_{\bf p q} V_0+g_{p^2 {\bf pq}} V_4&g_{({\bf p q})^2} V_1&g_{\bf p q} V_2\cr
g_\epsilon V_0 ++g_{p^2 \epsilon} V_4 &g_{\epsilon {\bf pq}} V_1 & g_\epsilon V_2}\right )_{\omega+{i \over \tau}}
\nonumber\\
{\cal B}&=&-\left (\matrix{d_1&e_1 & f_1\cr d_{\bf pq} & e_{\bf pq}& f_{\bf pq}\cr d_\epsilon & e_\epsilon & f_\epsilon} \right )
\nonumber\\
{\cal A}&=&-\left (\matrix{a_1&b_1 & c_1\cr a_{\bf pq} & b_{\bf pq}& c_{\bf pq}\cr a_\epsilon & b_\epsilon & c_\epsilon} \right )
\ee
and the abbreviations are introduced
\be
d_{\phi}&=&{-i\over \omega \tau+i} \left [ g_{\phi}(\omega+{i\over \tau})-g_\phi(0)\right ],\nonumber\\
e_{\phi}&=&{-i\over \omega \tau+i} \left [ g_{\phi {\bf q} \partial_{\bf p} \epsilon}(\omega+{i\over \tau})-g_{\phi {\bf q} \partial_{\bf p} \epsilon} (0)\right ],\nonumber\\
f_{\phi}&=&{-i\over \omega \tau+i} {1\over T} \left ( g_{\phi \epsilon}(\omega+{i\over \tau})-g_{\phi \epsilon}(0) \right .
\nonumber\\&&\left .- \mu \left [ g_{\phi}(\omega+{i\over \tau})-g_\phi(0)\right ]\right )\nonumber\\
\ee
in terms of the correlation function (\ref{g}).
The required solution is obtained from (\ref{g1}) and (\ref{g2}) as
\be
\left (\matrix{\delta n\cr\delta J_q\cr \delta E} \right )=({\cal I}-{\cal V}-{\cal B}{\cal A}^{-1})^{-1}\left (\matrix{g_1\cr g_{\bf pq}\cr g_\epsilon}\right ) V^{\rm ext}
\label{sol}
\ee
from which one can read off the response functions (\ref{def}).
This is the main result of the paper which represents the density, momentum and energy response including nonlinear mean fields and collisions with the fulfillment of
density -, momentum -, and energy - conservation.

\subsection{Alternative expressions}

Before we continue to consider special cases we like to express the solution (\ref{sol}) in a slightly more familiar form.

\subsubsection{Response in terms of mean-field response}

First we assume that we have solved the response without collisions ${\cal B}=0$ which would obey the equation
\be
({\cal I}-{\cal V}) \xi_0 ={\cal G} \nu^{ext}
\label{mfv}
\ee
where $\xi={\{}\delta n,\delta J_q,\delta E {\}}$, 
$\nu^{ext}={\{} V^{\rm ext},0,0 {\}}$ and

\be
{\cal G}(\omega)=\left (
\matrix {
g_1         &g_{\bf pq}            & g_\epsilon         \cr
g_{\bf p q} &g_{({\bf p q})^2}     & g_{{\bf p q}\epsilon} \cr
g_\epsilon  &g_{\epsilon {\bf pq}} & g_{\epsilon \epsilon} 
}
\right ).
\ee
Than the response matrix (\ref{def}) without collisions but selfconsistent mean field reads
\be
{\cal X}_{\rm MF}(\omega)=(1-{\cal V})^{-1} {\cal G}(\omega).
\ee
The missing part of the full solution of (\ref{g1},\ref{g2}) including the collisions are given by $\xi=\xi_0+\zeta$ where we have for $\zeta$
\be
(1-{\cal V}-{\cal B A}^{-1}) \zeta={\cal B A}^{-1} \xi_0.
\ee
Some algebra leads to the final response
\be
&&{\cal X}(\omega)=
\nonumber\\
&&{\cal X}_{\rm MF}(\omega+{i\over \tau}) \left (1-{\cal G}^{-1}(\omega+{i\over \tau}){\cal B A}^{-1} {\cal X}_{\rm MF}(\omega+{i\over \tau})\right )^{-1}.
\nonumber\\
&&
\label{sol1}
\ee

\subsubsection{Response in terms of polarization function}
The opposite case is the usual way where we first solve the equation without selfconsistency by the mean field. This leads to the polarization function ${\cal P}=\{\Pi,\Pi_J,\Pi_E\}$ which we use to represent the response function which includes selfconsistency. Without mean field we have from (\ref{sol})
\be
({\cal I}-{\cal B A}^{-1}) \xi_0={\cal G} \nu^{ext}
\label{merm}
\ee
leading to the polarization function
\be
{\cal P}(\omega)=(1-{\cal B A}^{-1})^{-1} {\cal G}(\omega+{i\over \tau}).
\ee
The response function can be represented analogously to (\ref{sol1})
\be
&&{\cal X}(\omega)=
{\cal P}(\omega)\left \{{\cal I}-{\cal G}^{-1}(\omega+{i\over \tau}) \, {\cal V} \, {\cal P}(\omega) \right \}^{-1}.
\nonumber\\
&&
\label{pmf}
\ee
The generalization of the usual form $\chi=\Pi/(1-V\Pi)$ for simple mean fields can be recognized.

\section{Calculation of response functions}

In the following we consider some frequently occurring situations. Therefore 
we assume only quadratic dispersions $\varepsilon=p^2/2m$ 
in the correlation function. To consider the full quasiparticle (Skyrme) energy would correspond to the selfconsistent quasiparticle RPA which we do not want to consider here. The schema how to include this is clear after the following considerations. We understand in the following the mass as effective mass given by $\epsilon_1$ in (\ref{pa}).  

The different occurring correlation functions (\ref{g}) can be written in terms of moments of the usual Lindhard polarization function $\Pi_0$
\be
\Pi_n=\int {d {\bf p}\over (2 \pi)^3} p^n{f_0({\bf p}+{{\bf q}\over 2})-f_0({\bf p}-{{\bf q}\over 2}) \over {{\bf p q}\over m}-\omega -i0}
\label{pn}
\ee
as following:
\be
g_1&=&\Pi_0
\nonumber\\
g_{\bf pq}&=&m \omega \Pi_0
\nonumber\\
g_\epsilon&=&{\Pi_2\over 2 m}
\nonumber\\
g_{\epsilon p^2}&=&{\Pi_4 \over 2m}
\nonumber\\
g_{p^2 {\bf pq}}&=&m \omega \Pi_2
\nonumber\\
g_{({\bf pq})^2}&=&-2 m q^2 {n\over s} + m^2 \omega^2 \Pi_0.
\label{gg}
\ee

For practical and numerical calculations we can rewrite the $\Pi_n$ by 
polynomial division into
\be
\Pi_2&=&-m {n\over s} +{m^2 \omega^2 \over q^2} \Pi_0 +\tilde \Pi_2
\nonumber\\
\Pi_4&=&-{14\over 3} m^2 {E_0\over s} -{n m q^2 \over 4 s} (1+{4 m^2 \omega^2 \over q^4})-{m^4 \omega^4 \over q^4} \tilde \Pi_0
\nonumber\\
&&-{2 m^2 \omega^2 \over q^2} \tilde \Pi_2-\tilde\Pi_4
\ee
where the $\tilde \Pi_i$ are the projected moments perpendicular to ${\bf q}$
and read
\be
\tilde \Pi_2&=&\int {d {\bf p}\over (2 \pi)^3} ({\bf p}-{{\bf pq}\over q^2}{\bf q} )^2 {f_0({\bf p}+{{\bf q}\over 2})-f_0({\bf p}-{{\bf q}\over 2}) \over {{\bf p q}\over m}-\omega-i0 }\nonumber\\
&=&2 m \int \limits_{-\infty}^\mu d\mu' \Pi_0
\nonumber\\ 
&\approx&2 m T \Pi_0
\nonumber\\
\tilde \Pi_4&=&\int {d {\bf p}\over (2 \pi)^3} ({\bf p}-{{\bf pq}\over q^2}{\bf q} )^4 {f_0({\bf p}+{{\bf q}\over 2})-f_0({\bf p}-{{\bf q}\over 2}) \over {{\bf p q}\over m}-\omega-i0 }\nonumber\\
&=&2 (2 m)^2 \int \limits_{-\infty}^\mu d\mu'\int \limits_{-\infty}^{\mu'} d\mu'' \Pi_0 
\nonumber\\
&\approx&8 m^2 T^2 \Pi_0.
\label{ein}
\ee
The corresponding last identities are valid only for nondegenerate, Maxwellian, distributions with temperature $T$. The general form of polarization functions is presented as an integral over the chemical potential $\mu$ of the Lindhard polarization $\Pi_0$. This is applicable also to the degenerate case.
In the following we will discuss successively further involved results; first for nondegenerate plasmas and then for degenerate nuclear matter.

\subsection{Polarization with collisions: inclusion of density and energy conservation}

Now we concentrate on the response function without mean field and consider only
the collisions within density and energy conservation. Then the matrices
${\cal A}$ and ${\cal B}$ reduces to 2x2 matrices. The calculation of (\ref{merm}) leads to
\be
\Pi^{\rm n,E}(\omega)&=&(1-i \omega \tau)\left ({g_1(\omega+{i\over \tau}) g_1(0)\over h_1}
\right .\nonumber\\
&&\left . -\omega\tau i {(h_\epsilon g_1(0)-h_1 g_\epsilon(0))^2\over h_1(h_\epsilon^2-h_{\epsilon \epsilon} h_1)}\right )
\label{m-ne}
\ee
where we use the abbreviation
\be
h_\phi=g_\phi(\omega+{i\over \tau})-\omega \, \tau \, i \, g_\phi(0).
\label{h}
\ee
With the help of (\ref{gg}-\ref{ein}) this can be further worked out in terms of $\Pi_n$ but leads not to a more transparent form.
Let us note that the first term in (\ref{m-ne}) represents just the result if we would have considered only density conservation known as the Mermin polarization function \cite{Mer70}
\be
\Pi^{\rm n}(\omega)&=&(1-i \omega \tau){g_1(\omega+{i\over \tau}) g_1(0)\over h_1}.
\label{m-n}
\ee

Of course the limit of vanishing collisions $\tau\rightarrow \infty$ ensures that the Lindhard result $\Pi_0(\omega)$ appears since
\be
\lim\limits_{\tau \to \infty} {h_\phi\over 1-i \omega \tau}=g_\phi(0).
\ee

\begin{figure}
\centerline{\psfig{file=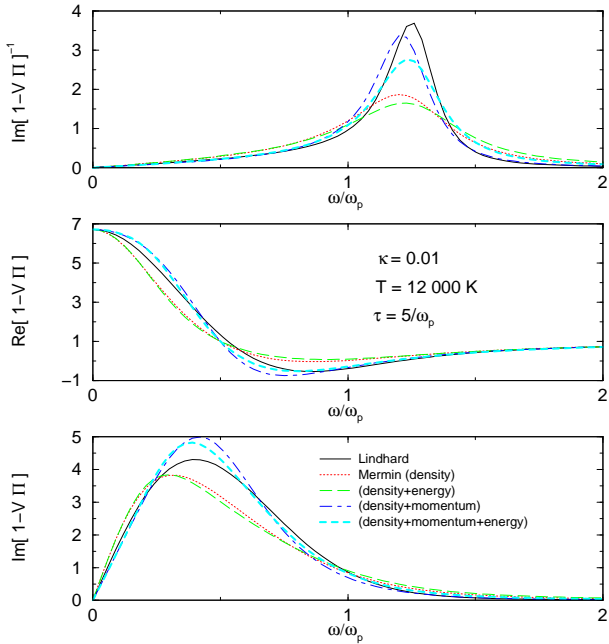,width=8cm,angle=-90}}
\caption{The dielectric function $\varepsilon=1-V \Pi$ for a one component plasma system in different approximations. The inverse Debye length is $\kappa^2={4 \pi n e^2\over T}$ and the energy $\omega$ is scaled in plasma frequencies $\omega_p^2=\kappa^2 T/m$. The relaxation time is chosen arbitrarily as $\tau=5/\omega_p$ and the wave vector $q=0.384 \kappa$. The upper most panel shows the excitation function.}

                      \label{ff1}
\end{figure}

\subsection{Polarization with collisions: inclusion of density and momentum conservation}

Next we consider the special case that
the density and momentum are conserved. Then the matrices
${\cal A}$ and ${\cal B}$ reduces again to 2x2 matrices and the calculation of (\ref{merm}) leads to
\be
\Pi^{\rm n,J}(\omega)&=&(1-i\omega \tau){g_1(\omega +{i\over \tau })-{\left (g_{\bf pq}(\omega +{i\over \tau })\right )^2\over h_{({\bf pq})^2}}\over h_1-{\left (g_{\bf pq}(\omega +{i\over \tau })\right )^2\over h_{({\bf pq})^2}}} g_1(0).
\label{m-np}
\ee
We have used the fact that according to (\ref{gg}) $g_{\bf pq}(0)=g_{\epsilon{\bf pq}}(0)=0$ and $h_\phi$ is defined as in (\ref{h}). With the help of (\ref{eps}) we obtain by this way a slightly modified Mermin dielectric function (\ref{m-n}).

\subsection{Polarization with collisions: inclusion of density, momentum and energy conservation}

Considering all three conservation laws the result from (\ref{pmf}) is 
\be
&&\Pi^{\rm n,J,E}(\omega)=(i \omega \tau-1) \left (i\omega \tau {N\over D}-g_1[0]\right ),
\label{m-npe}
\ee
with
\be
N=&&-\left \{g_{\epsilon {\bf pq}}[\omega+{i\over \tau}] g_1[0] 
-g_{\epsilon}[0] g_{\bf pq}[\omega+{i\over \tau}]\right \}^2 
\nonumber\\&&
+
h_{({\bf pq})^2} \left \{  g_1[0] \left (h_{\epsilon^2} g_1[0] -h_{\epsilon}g_{\epsilon}[0]\right )\right .\nonumber\\
&&\left . \quad \qquad +
g_{\epsilon}[0]\left (h_1 g_{\epsilon}[0]-h_{\epsilon} g_1[0]\right )\right \}
\ee
\be
D&=&
g_{\epsilon {\bf pq}}[\omega+{i\over \tau}]\left \{h_1 g_{\epsilon {\bf pq}}[\omega+{i\over \tau}]-
h_{\epsilon} g_{\bf pq}[\omega+{i\over \tau}] \right \}
\nonumber\\&+&
g_{\bf pq}[\omega+{i\over \tau}] \left \{h_{\epsilon \epsilon}g_{\bf pq}[\omega+{i\over \tau}]- h_{\epsilon}g_{\epsilon {\bf pq}}[\omega+{i\over \tau}]\right \}
\nonumber\\&+&
h_{({\bf pq})^2}\left \{h_{\epsilon}^2-h_{\epsilon \epsilon} h_1\right \}.
\ee
This result together with the former special cases
(\ref{m-n},\ref{m-ne},\ref{m-np}) are compared in figure
\ref{ff1}. One sees that the first approximation of Mermin (\ref{m-n})
is almost identical with the result (\ref{m-ne}) where density and
energy is conserved. The inclusion of density and momentum
conservation (\ref{m-np}) brings the curves towards the Lindhard
result without collisions compared with the inclusion of density
conservation only. Finally, the complete result with the inclusion of density, momentum and energy conservation (\ref{m-npe}) changes the results again in the direction of the result for density and energy conservation but less pronounced. This qualitative behavior of the different approximations are observed for other relaxation times too.

\begin{figure}
\centerline{\psfig{file=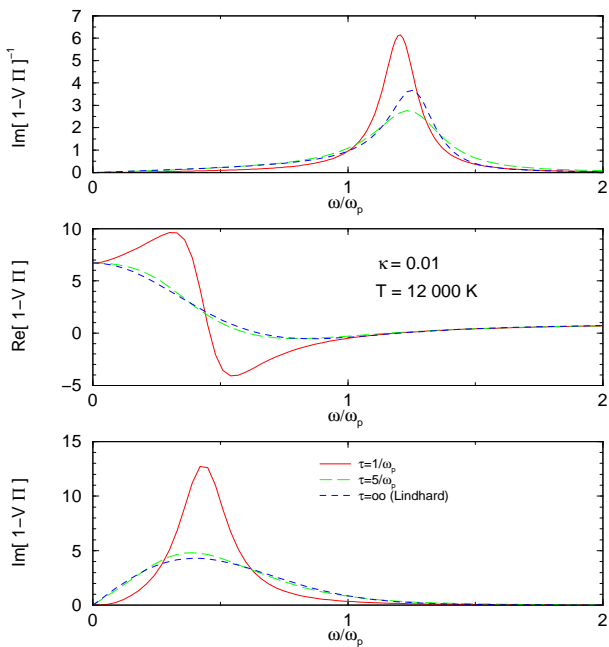,width=8cm,angle=-90}}
\caption{The dielectric function $\varepsilon=1-V \Pi$ for a one component plasma system with the inclusion of density, momentum and energy conservation (\protect\ref{m-npe}) for different relaxation times. The parameter are the same as in figure \protect\ref{ff1}.}
                      \label{ff2}
\end{figure}

The effect of relaxation times within the complete result (\ref{m-npe}) is seen in figure \ref{ff2}. One recognizes that with decreasing relaxation time or increasing collision frequency the plasmon peak is shifted towards smaller energies. For collision frequencies around the inverse plasma frequency there occurs a resonance seen in the real part of the dielectric function (middle part of \ref{ff2}). This translates into an enhanced single particle damping (lower panel) and the system becomes optical thick. At the same time the collective mode, the plasma frequency, becomes enhanced. One can consider this as an effect of transferring collisional energy into collective motion. We have here a coherent superposition between collision frequency and collective frequency resulting into an enhancement of collective motion. Compared with the general effect of collisions to increase the damping of collective motion, see figure \ref{ff1}, this is the inverse effect which narrows the collective peak again.

\subsection{Response with collisions: simple mean field}

For the case of simple but density - dependent mean fields $V_0\ne0$ and $V_1=V_2=V_4=0$ we obtain for the density response from (\ref{pmf})
\be
\chi={\Pi\over 1-V_0 \Pi}.
\label{sm}
\ee
Here $\Pi$ is the polarization function without mean field but with collisions. Dependent on the choice one may use (\ref{m-ne}), (\ref{m-np}) or (\ref{m-npe}) for the latter one.

\begin{figure}
\centerline{\psfig{file=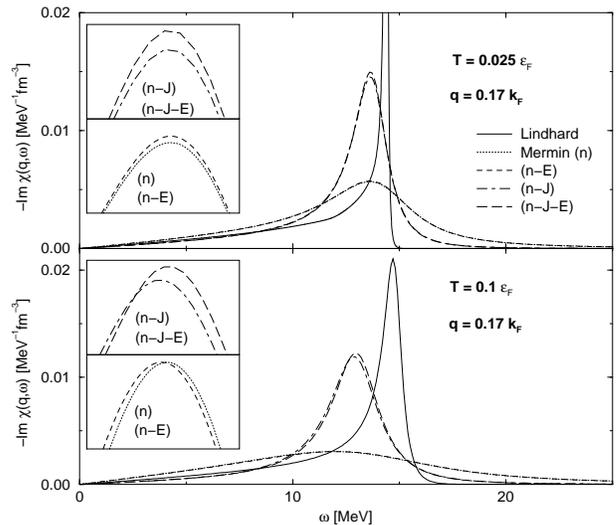,width=8cm,angle=-90}}
\caption{The imaginary part of the response function ${\rm Im}\chi$ for 
nuclear giant dipole resonances in different approximations 
(comp. Fig. \protect\ref{ff1}). The wave vector 
$q= 0.23\,{\rm fm}^{-1}$ ($\approx 0.17\,{\rm k}_F$) corresponds 
to the inverse diameter of the nucleus $^{208}$Pb 
according to \protect\cite{SJE50}. The inlays
show an enlarged view of the difference between the density (n),density-
energy (n-E) and density-momentum (n-p), density-momentum-energy (n-p-E)
approximation, respectively. }
\label{struk024T1T4V0} 
\end{figure}

Since the imaginary part of the response function (\ref{sm}) is related to the
photoabsorbtion yield on nuclei we like to apply the different approximations
(\ref{m-ne}), (\ref{m-np}) or (\ref{m-npe}) for nuclear isovector oscillations.
We use first a simplified Skyrme parameterization \cite{BV94} 
for $V_0$ according to (\ref{skyrmeV0V1}) and a relaxation time $\tau$ within
a Fermi liquid model \cite{FMW98,MFU99}.
From Fig. \ref{struk024T1T4V0} we see the same qualitative behavior of the
different approximations as found in Fig. \ref{ff1}. The difference
between the density (density-momentum)  and density-energy (density-energy-
momentum) result is very small (inlays of Fig. \ref{struk024T1T4V0}).
An increase of temperature leads to larger damping of all approximations 
(bottom of Fig. \ref{struk024T1T4V0}).  While the density or density-energy result leads to a pronounced damping of the giant resonance the inclusion of momentum conservation diminishes this effect again towards the free result.

Let us note that the energy-weighted sum rule (EWSR)
\be
-\frac{1}{\pi}\int_0^{\infty}d\omega\, \omega\,{\rm Im}\chi=\frac{q^2}{2 m} n_0
\label{fsum}
\ee
is fulfilled numerically for all approximations, 
however, the convergence is very bad for response including density or 
density and energy conservation. The inclusion of momentum conservation in turn 
improves the convergence of the sum rule appreciable.

\subsection{Response without collisions: Skyrme mean field}
First we consider the case that we have only Skyrme - mean fields. Then the 
matrix equation (\ref{mfv}) is solved with the result for the density response function
\be
&&\chi_{\rm MF}(\omega,V_i)=\nonumber\\
&&{\Pi_0\over 1-\Pi_0 {\bar V_0} + {V_2 V_4 \over 2 m} \left [
\Pi_2^2 -\Pi_0 \Pi_4  \right ]-\Pi_2 \left [{V_2 \over 2 m}+V_4\right ]}\nonumber\\
\label{mfr}
\ee
with
\be
{\bar V_0}=V_0+{m \omega ( m \omega V_1+V_3)\over 2 {n\over s} m q^2 V_1 +1}.
\ee 
For isovectorial oscillations one has
\be
V_0&=&{\delta \varepsilon_0 \over \delta n}=V_0^s-V_1^s{q^2\over 2}\nonumber\\
V_1&=& {\delta \varepsilon_0 \over \delta {J_q}}=-{2\over q^2} V_1^s\nonumber\\
V_2&=&{\delta \varepsilon_0 \over \delta E}=2 m V_1^s\nonumber\\
V_4&=&V_1^s
\label{mf1}
\ee
where the $V_i^s$ are representing the Skyrme parameterization in nuclear matter \cite{BVA95} 
\be
V_0^s&=&-t_0 (x_0+\frac 1 2)-{t_3 \over 6} (x_3+\frac 1 2)n^\alpha_0 
\nonumber\\&&-{q^2\over 16}[3 t_1 (1+2 x_1)+t_2 (1+ 2 x_2)]\nonumber\\
V_1^s&=&{1\over 8}[t_2(1+2 x_2)-t_1 (1+ 2 x_1)]. \label{skyrmeV0V1}
\ee
If we use the definitions of \cite{BVA95} which are related to ours as
\be
\tilde \Pi_0&=&\Pi_0\nonumber\\
\tilde \Pi_2&=&\Pi_2-{q^2\over 4} \Pi_0\nonumber\\
\tilde \Pi_4&=&\Pi_4-{q^2\over 2} \Pi_2+{q^4\over 16} \Pi_0
\ee
we obtain the result of \cite{BVA95} which was slightly misprinted
\be
\chi_{\rm MF}=
{\Pi_0\over 1-\tilde \Pi_0 \tilde V_0^s -2 V_1^s \tilde \Pi_2 +{(V_1^s)^2}\left [
\tilde \Pi_2^2-\tilde \Pi_0 \tilde \Pi_4\right ]}
\ee
with 
\be
\tilde V_0^s=V_0^s-{m^2 \omega^2\over q^2 }{2 V_1^s \over 1-n_0 m V_1^s}
\ee
for nuclear matter density $n_0$ and $s=4$.

\subsection{Response with collisions: inclusion of density
  conservation and Skyrme mean field}

Now we derive the combined result from the mean field response (\ref{mfr}) and collisions.
We restrict first only on density balance conservation [first term of 
(\ref{m-ne})] to get from (\ref{6})
\be
\delta \mu=-{\delta n\over a_1}=-{\delta J_q\over a_{\bf pq}}=-{\delta E\over a_\epsilon}
\ee
and in (\ref{sol})
\be
{\cal B A}^{-1}=\left (\matrix{{d_1\over a_1}&0&0\cr0&{d_{\bf pq}\over a_{\bf pq}}&0\cr0&0&{d_\epsilon\over a_\epsilon}}\right ).
\label{49}
\ee
Therefore we can solve (\ref{pmf}) and obtain similar to (\ref{mfr})
\be
&&\chi^{\rm n}(\omega)=\nonumber\\
&&(1-i\omega\tau){g_1(0)\over h_1} \chi_{\rm MF}(\omega+{i\over \tau},\tilde V_i)
\label{mfr1}
\ee
with
\be
\tilde V_0&=&(1-i\omega\tau){ g_1(0)\over h_1} V_0
\nonumber\\
\tilde V_1&=&(1-i\omega\tau){ g_{\bf pq}(0)\over h_{\bf pq}} V_1=\left \{
\matrix{V_1+o(\omega \tau)^{-1}\cr0 +o(\omega \tau)}\right .
\nonumber\\
\tilde V_2&=&(1-i\omega\tau){g_\epsilon(0)\over h_\epsilon} V_2
\nonumber\\
\tilde V_4&=&(1-i\omega\tau){ g_1(0)\over h_1} V_4.
\label{V}
\ee
We should note that the mean field potential $V_1$ arising from current and effective mass is of a different level of approximation than the collisional contribution which we restricted in (\ref{49}) to the density conservation. This inconsistency is visible in the final result for $\tilde V_1$ in (\ref{V}) where the limit of vanishing collisions $\tau \to \infty$ does not agree with the limit of finite collisions since $g_{\bf pq}(0)\equiv0$, see (\ref{gg}).
Therefore $\tilde V_1=V$ is proposed to ensure the limit of vanishing collisions.

\begin{figure}
\centerline{\psfig{file=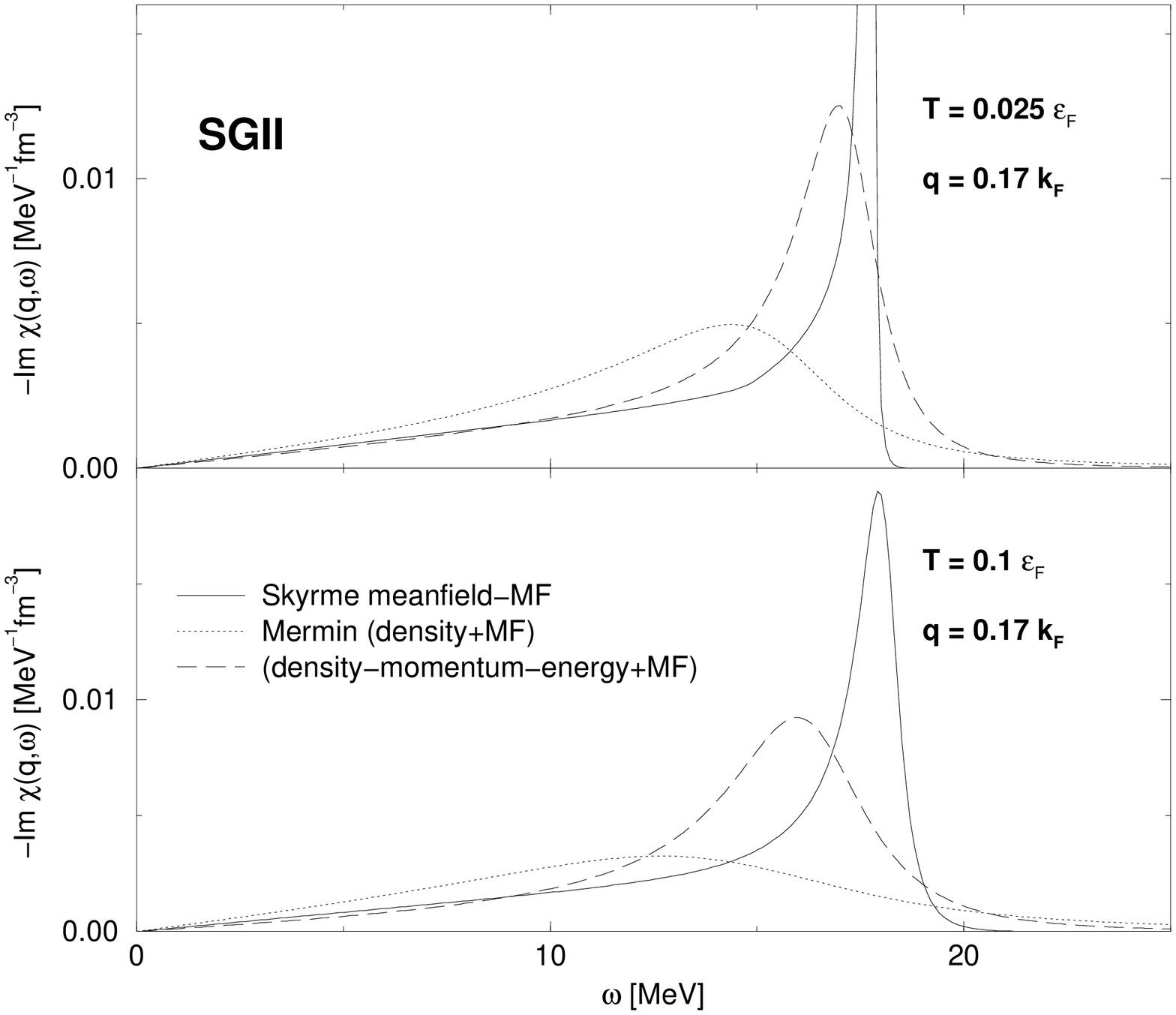,width=7.5cm,angle=-90}}
\caption{The imaginary part of the response function ${\rm Im}\chi$ for 
nuclear giant dipole resonances (see Fig. \protect\ref{struk024T1T4MF})
in different approximations. Considering the Skyrme 
interactions SGII \protect\cite{BVA95,GSA81} we compare the full mean field 
response  with collisional correlations in density (dotted line) and density, momentum and 
energy  approximation (dashed line), respectively, with the collision-free
mean field response (solid line) for two different temperatures. }
\label{struk024T1T4MF} 
\end{figure}

\subsection{Response with collisions: inclusion of density, momentum and 
energy conservation and Skyrme mean field}

The Skyrme response can be given also for the other cases 
including energy and momentum conservation. However this leads not to a more 
transparent form than the general matrix structure (\ref{pmf}). 
Considering the standard effective Skyrme interaction SGII \cite{GSA81}
in figure \ref{struk024T1T4MF} we compare the complete result (\ref{sol}) including 
energy, momentum and density conservation (dashed line) with the result including only 
density conservation (\ref{mfr1}) (dotted line). 
In the latter one we have proposed for $\tilde V_1$ the collision-free 
value $V_1$ in order to ensure the correct limiting case.

We find again the same behavior for the different approximations 
as in the case of the simplified mean field (Fig. \ref{struk024T1T4V0}). 
The inclusion of collisions leads to an enhanced damping and a shift of the collective peak towards smaller energies.
This effect of collisions is less pronounced
by the full mean field result  including density, momentum and
energy conservation (solid line).  The increase of temperature leads to a broadening
of the resonance structure (lower part of Fig. \ref{struk024T1T4V0}) in any case.

Again we have checked our results to satisfy the EWSR
\be
-\frac{1}{\pi}\int_0^{\infty}d\omega\, \omega\,{\rm Im}\chi=\frac{q^2}{2 m} 
n_0 (1+\kappa), \label{fsumM}
\ee
with the enhancement factor $\kappa$ 
\be 
\kappa=\frac{m}{4}\left[t_1 (2+x_1)+t_2 (2+x_2)\right]n_0.
\ee
Here $\kappa$ occurs as a consequence 
of the momentum-dependent terms in the Skyrme interaction
and is defined as the deviation from the 
Thomas-Reiche-Kuhn sum rule in the case of
isovector giant dipole resonance \cite{MQJ82,GBM90}.

The result (\ref{mfr}) which contains the full Skyrme mean field
but no collisions
is in excellent agreement with (\ref{fsumM}). The approximation (\ref{mfr1})
including collisions but only density conservation conserves the sum rule 
only $\approx 75\%$ which is a consequence of the inconsistency of (\ref{V}).
The inclusion of density, momentum and energy conservation (\ref{sol}) 
conserves the sum rule (\ref{fsumM}) again completely.

\section{Summary}

In this paper we have derived the unifying response function including
nonlinear mean fields (Skyrme) and collisional correlations. Within
this approach one can share correlations between an energy
functional of mean field  (Skyrme) parameterization and explicit
dissipative correlations condensed in the relaxation time. This allows
to treat also dissipative effects in density functional approaches. 
We see that the known limiting cases are reproduced neglecting either collisions or mean fields. Special transparent cases of the unifying response are discussed. 

For a nondegenerate plasma numerical results are presented. The first
order correction given by the Mermin response, incorporating only the
density balance, is similar to the approximation where density and
energy are conserved. The plasmon peak is shifted towards smaller
frequencies. This is accompanied by an enhanced damping. The
incorporation of momentum balance diminishes this effect of
collisions. 

We observe that an enhancement of the collective mode occurs for collision frequencies near the collective (plasma) frequency. This is the inverse effect of damping due to collisions in that the collisions become resonant and the collective mode is enhanced. We consider this as collisional narrowing. Since the momentum conservation is responsible for that effect we suggest that the physical origin is the same as sometimes discussed with motional narrowing. We like to stress that this narrowing is observed relative to the broadened mode due to collisions and did not reach the collision - free value. Consequently we have collisional damping every time but near the resonant situation this collisional damping is diminished.     

Similar behavior is found for the case of nuclear matter, where the collective mode is the giant resonance. For isovectorial giant resonances we checked the extended energy weighted sum rules and find excellent completion. The response due to nonlocal mean fields is derived including the effect of collisional correlations. 

\acknowledgements
The authors like to thank H.S. K{\"o}hler for discussions and helpful comments.


\begin{thebibliography}{10}

\bibitem{BVA95}
F. Braghin, D. Vautherin, and A. Abada, Phys. Rev. C {\bf 52},  2504  (1995).

\bibitem{HNP96}
E. Hern\'andez, J. Navarro, A.Polls, and J. Ventura, Nucl. Phys. {\bf A597},  1
   (1996).

\bibitem{Mer70}
N. Mermin, Phys. Rev. B {\bf 1},  2362  (1970).

\bibitem{HPR93}
H. Heiselberg, C.~J. Pethick, and D.~G. Ravenhall, Ann. Phys. {\bf 223},  37
  (1993).

\bibitem{KH94}
D. Kiderlen and H. Hofmann, Phys. Lett. B {\bf 332},  8  (1994).

\bibitem{RW98}
G. R{\"o}pke and A. Wierling, Phys. Rev. E {\bf 57},  7075  (1998).

\bibitem{MWF97}
K. Morawetz, R. Walke, and U. Fuhrmann, Phys. Rev. C {\bf 57},  {R 2813}
  (1998).

\bibitem{Br99}
F. Braghin, Phys. Lett. B {\bf 446},  1  (1999).

\bibitem{S56}
T.~H.~R. Skyrme, Phil. Mag. {\bf 1},  1043  (1956).

\bibitem{S59}
T.~H.~R. Skyrme, Nucl. Phys. {\bf 9},  615  (1959).

\bibitem{LSM99}
P. Lipavsk{\'y}, V. {\v S}pi{\v c}ka, and K. Morawetz, Phys. Rev. E {\bf 52},
  R1291  (1999).

\bibitem{EBGKV75}
Y. Engel {\it et~al.}, Nucl. Phys. A {\bf 249},  215  (1975).

\bibitem{Ko75}
H.~S. K{\"o}hler, Nucl. Phys. A {\bf 258},  301  (1976).

\bibitem{BV94}
F. Braghin and D. Vautherin, Phys. Lett. B {\bf 333},  289  (1994).

\bibitem{FMW98}
U. Fuhrmann, K. Morawetz, and R. Walke, Phys. Rev. C. {\bf 58},  1473  (1998).

\bibitem{MFU99}
K. Morawetz and U. Fuhrmann, in {\em Proceedings of the Riken Symposium on
  Nuclear Collective Excitations},Saitama, March 1999.

\bibitem{SJE50}
H. Steinwedel and J. Jensen, Z. Naturforsch. {\bf 5},  413  (1950).

\bibitem{GSA81}
N.~V. Giai and N. Sagawa, Nucl. Phys. A {\bf 371},  1  (1981).

\bibitem{MQJ82}
P.~Q. J.~Meyer and B.~K. Jennings, Nucl. Phys. A {\bf 385},  103  (1982).

\bibitem{GBM90}
J.~M. P.~Gleissl, M.~Brack and P. Quentin, Ann. Phys. (NY) {\bf 197},  205
  (1990).

\end{thebibliography}

\end{document}